\documentclass[12pt,english]{article}
\usepackage[T1]{fontenc}
\usepackage[latin9]{inputenc}
\usepackage{geometry}
\usepackage{color}
\usepackage{bm}
\usepackage{amstext}
\usepackage{graphicx}
\usepackage{setspace}
\usepackage{esint}
\usepackage[numbers]{natbib}
\makeatletter
\usepackage{blindtext}

\makeatother

\usepackage{babel}
\begin{document}

\title{Helical vortices generated by flapping wings of bumblebees}

\author{Thomas Engels$^{1,4}$, Dmitry Kolomenskiy$^{2}$, Kai Schneider$^{3}$,
Marie Farge$^{4}$, \\
Fritz-Olaf Lehmann$^{5}$ and Jörn Sesterhenn$^{1}$\\
{\footnotesize{}$^{1}$ ISTA, Technische Universität Berlin, Berlin,
}\\
{\footnotesize{} Müller-Breslau-Strasse 12, 10623 Berlin, Germany
}\\
{\footnotesize{} $^{2}$ CEIST, Japan Agency for Marine-Earth Science
and Technology (JAMSTEC). }\\
{\footnotesize{}3173-25 Showa-machi, Kanazawa-ku, Yokohama Kanagawa
236-0001, Japan. }\\
{\footnotesize{} $^{3}$ CNRS, Centrale Marseille, I2M, Aix-Marseille
Université, Marseille, }\\
{\footnotesize{} 39 rue Joliot-Curie, 13451 Marseille Cedex 20 France
}\\
{\footnotesize{} $^{4}$ LMD-CNRS, Ecole Normale Supérieure and PSL,
Paris, }\\
{\footnotesize{} 24 rue Lhomond, 75231 Paris Cedex 05, France}\\
{\footnotesize{} $^{5}$ Department of Animal Physiology, Universität
Rostock, Rostock, }\\
{\footnotesize{} Albert-Einstein-Str. 3, 18059 Rostock, Germany}}

\maketitle
\begin{abstract}
High resolution direct numerical simulations of rotating and flapping
bumblebee wings are presented and their aerodynamics is studied focusing
on the role of leading edge vortices and the associated helicity production.
We first study the flow generated by only one rotating bumblebee wing
in circular motion with $45^{\circ}$ angle of attack. We then consider
a model bumblebee flying in a numerical wind tunnel, which is tethered
and has rigid wings flapping with a prescribed generic motion. The
inflow condition of the wind varies from laminar to strongly turbulent
regimes. Massively parallel simulations show that inflow turbulence
does not significantly alter the wings' leading edge vortex (LEV),
which enhances lift production. Finally, we focus on studying the
helicity of the generated vortices and analyze their contribution
at different scales using orthogonal wavelets. 
\end{abstract}

\section{Introduction\label{sec:Introduction}}

Numerical modeling of flapping insect flight receives considerable
attention and is motivated by the growing interest in miniaturization
of unmanned air vehicles, since flapping wings present a bio-inspired
alternative to the fixed- and rotary-wings used in human-designed
aircraft. The force production in those two kind of fliers relies
on entirely different aerodynamic mechanisms. Airplane wings are smooth
and use airfoil shapes designed to produce lift from an attached flow
which is accelerated on the suction side. Flow separation (stall)
limits the range of angle of attack in which these airfoils are useful.
By contrast, insect wings feature sharp edges, essentially flat profile
and large angles of attack. Under these conditions, flow separation
is inevitable and large amounts of vorticity are generated at the
leading edge. This vorticity forms a strong vortex which moves with
the wing and detaches only at the stroke reversals. It has been suggested
that insects can capture it at early times in the following half-stroke
to provide an additional benefit \citep{Lehmann2007}. Some insects
clap their wings together and the subsequent opening motion creates
a fluid jet which also provides additional forces. This mechanism
is known as clap-fling-sweep \citep{weisfogh1973} and it has recently
been revisited \citep{KMFS11b}. Dragonflies and some other species
can control their four wings independently and have arranged them
in a configuration that allows aerodynamic interaction between fore-
and hindwing. This interaction depends on the phase difference in
their kinematics and can contribute to force production as well \citep{UsherwoodLehmann2008},
\citep{Kolomenskiy2013}, \citep{Inamuro2015}.

Previous research on the flow generated by flapping wings indicates
the important role of the leading edge vortex (LEV) \citep{Liu_etal_1998_jeb,Ellington_etal_1996_nature}.
This vortex has a conical structure due to the three-dimensional motion
of the wings. Vorticity is produced at the sharp leading edge, and
outwards velocity (from the root to the tip of the wing) develops
above the suction surface of the wing, see, e.g., \citep{Maxworthy_1979_jfm,KMFS11,Kolomenskiy2014a}.
Such alignment of the vorticity and the velocity has important consequences
for the dynamics of the vortex \citep{Chen2017}. On one hand, the
excess vorticity is constantly transported into the wing tip vortex
rather than being shed periodically from the leading edge \citep{Maxworthy_1979_jfm}.
On the other hand, the swirl angle is large and the vortex can burst
\citep{Maxw2007}. Swirling flows are characterized by strong helicity,
which is defined by the scalar product of velocity and vorticity vectors
and corresponds to their alignment or anti-alignment. Consideration
of the helicity dynamics in flows over flapping or revolving wings
can therefore bring important insights into the processes that determine
the flow topology. 

Helicity has received much attention in the topological fluid dynamics
community to measure the linkage and knottedness of vortex lines in
the flow. For a review we refer for instance to \citep{MoTs92}. In
the turbulence community helicity has been used to characterize three-dimensional
swirling coherent structures, which correspond to flow regions of
maximum helicity \citep{FaPS01}. This local alignment or anti-alignement
of velocity and vorticity implies that the nonlinear term of the Navier\textendash Stokes
equations is depleted and thus the nonlinear energy transfer is slowed
down. This energy cascade, also known as Kolmogorov cascade, transfers
energy from larger to smaller and smaller scales until it is eventually
dissipated. Its inhibition in regions of strong helicity indicates
that these structures tend to be more stable and to persist coherently
in time \citep{Moff14}. An example for flows with maximum helicity
are Beltrami flows, which correspond to eigenfunctions of the curl
operator and are hence solutions of the steady Euler equations.

To get insight into the scale distribution of helicity we decompose
the velocity and vorticity into orthogonal wavelet bases. Wavelets
are localized functions in scale and space and allow analyzing flow
fields efficiently. Thus the scale-dependent helicity, introduced
in \citep{YOSKF09}, can be computed. A review on wavelet based statistical
measures for fluid and plasma turbulence can be found in \citep{FaSc15}.

The aim of this work is to examine the helicity dynamics in flows
over model insect wings in connection with the effects that were previously
described in terms of the vorticity and the velocity. We propose helicity
as a new diagnostics to study the vortices generated by flapping and
revolving wings. Although it is often stated that the leading edge
vortex is 'helical', its helicity, in the sense of its proper mathematical
definition, has received surprisingly little attention. It has only
been used to discuss the bursting of the leading edge vortex on a
revolving wing, i.e. its transition from a simple to a more complex
topology \citep{Jones2016}. This transition is reflected as a drop
in the volume integral of helicity in the leading edge vortex, but
not in the generation of aerodynamic force, as such a burst vortex
still induces a locally reduced pressure. We should however stress
the difference between vortex bursting, i.e. the change from a simple
to a more complex topology, and vortex shedding. In the latter, the
leading edge vortex periodically leaves the vicinity of the wing and
constitutes a wake, and consequently the aerodynamic forces oscillate,
with a significantly reduced mean value \citep{Kolomenskiy2014a}.
Whether vortex shedding occurs or not depends, amongst others, on
the Rossby number and the wing aspect ratio \citep{lentink2008,Lee2016}.
In a different context, namely the wing/wing interaction of fore-
and hindwing in dragonflies, the 'swirl' of the wake has been discussed,
\citep{UsherwoodLehmann2008}, but swirl was therein considered as
measure for lateral impulse transport rather than the helicity as
considered here.

First, we investigate a simplified configuration of an unilaterally
rotating bumblebee wing and perform high resolution numerical computations.
The flow fields are studied and, in particular, the leading edge vortex
is examined. Second, we analyze data of a flapping bumblebee flying
in turbulent flow, presented in \citep{EKSLS15}. We use the orthogonal
wavelet decomposition of the flow field to analyze the production
of helicity at different scales, which is then quantified by the wavelet
spectrum of helicity and its spatial variability.

The manuscript is organized as follows: In section~2 we describe,
for reasons of self-consistency, the bumblebee model with rigid wings
and the computational set-up. The wing kinematics and parameters can
be found in the cited references. The numerical method, which is a
Fourier pseudo-spectral method with volume penalization, is briefly
recalled too. The computational results are reproducible as the ``FluSI''
code is open source \citep{EKSS15}. The definition of helicity, together
with its spectral decomposition and the scale-dependent helicity using
orthogonal wavelets are also given. Computational results for rotating
and flapping bumblebee wings are presented and subsequently analyzed
in section~3. Conclusions of our findings are drawn in section~4.

\section{Bumblebee model, numerical method and helicity\label{sec:materials-methods}}

\subsection{Bumblebee model\label{subsec:Bumblebee-model}}

In this article, a bumblebee (\textit{Bombus terrestris}), already
used in previous work \citep{EKSLS15}, is chosen among the variety
of flying insects as typical representative for medium-size species.
Bumblebees are known to be relentless all-weather foragers \citep{Wolf1999,Ravi2013,Crall2016}
and thus encounter a wide range of flow conditions from laminar to
fully turbulent \citep{EKSLS15,Crall2016}. The flow they generate
while flying remains in a range of Reynolds number which can be computed
by direct numerical simulation (DNS) using high-performance computing
facilities. The key parameters, which we use in both setups described
below, of the model insect are: wing length $R=13.2\,\mathrm{mm}$,
wingbeat frequency $f=152\,\mathrm{Hz}$, total mass $m=175\,\mathrm{mg}$,
forward flight speed $u_{\infty}=2.5\,\mathrm{m/s}$, Reynolds number
$Re=\overline{u}_{\mathrm{tip}}c_{m}/\nu=2060$, where $\overline{u}_{\mathrm{tip}}=8.05\,\mathrm{m/s}$
is the mean wingtip velocity, $c_{m}=A/R=4.01\,\mathrm{mm}$ the mean
chord length, $A=52.96\,\mathrm{mm^{2}}$ the wing surface and $\nu=15\cdot10^{-6}\,\mathrm{m^{2}/s}$
the viscosity of air. The planform of the wing is illustrated in Fig.
\ref{fig:setup}b. The wing is modeled as a rigid, flat surface.

A different definition of the Reynolds number can be based on the
mean velocity at the radius of gyration, $R_{2}=\sqrt{\int_{0}^{R}r^{2}c(r)\mathrm{d}r/A}=7.6032\,\mathrm{mm}$,
which yields $Re_{2}=1187$. Using $R_{2}$ has its root in the blade
element theory \citep{Elington1984}, where it appears naturally,
and it has been suggested to provide a better value for comparison
in the case of revolving wings, as it reflects also the aspect ratio
\citep{Lee2016}. Note that the velocity at the center of wing area
can also be used as reference velocity and it may be advisable for
the purpose of comparison between different flappers \citep{Lua2014}.
In this paper, however, we only consider one wing shape. As both Reynolds
numbers are common, we will use both definitions.

For the rest of the article we shall only use dimensionless quantities,
normalized with a length scale $L=R$, a mass scale $M=\varrho_{f}L^{3}$
(which implies that the dimensionless fluid density is unity) and
a time scale $T$, which we choose depending on the setup.

\subsubsection{One revolving wing: the canonical model \label{subsec:One-revolving-wing}}

\begin{figure}
\begin{centering}
\includegraphics[width=0.8\textwidth]{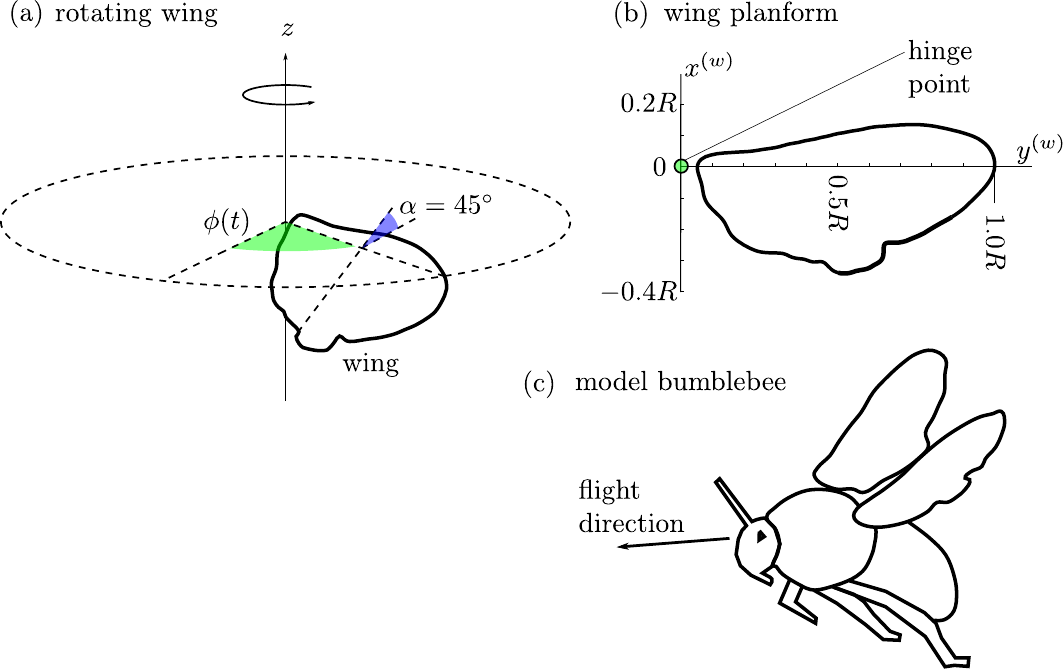}
\par\end{centering}
\caption{Setup: configuration for the revolving wing (a),wing mask (b) and
bumblebee mask (c). \label{fig:setup}}
\end{figure}

Prior to analyzing the complete insect model, we focus in this part
on a commonly used reduced model, which consists of a single, revolving
wing. This canonical setup is often used to study the leading edge
vortex \citep{Kolomenskiy2014a,Harbig2013,Garmann2014,Jones2016,Lee2016}.
We fix the angle of attack to $\alpha=45^{\circ}$ (i.e., the feathering
angle, for details see Figure~\ref{fig:setup} and \citep{EKSS15}).
The rotation angle varies as 
\[
\phi\left(t\right)=\dot{\Phi}\left(\tau e^{-t/\tau}+t\right),
\]
which is the same as used in previous work \citep{Kolomenskiy2014a}.
After a transient time, $\tau=0.4$, the rotation angle grows linearly
in time. The wingtip velocity is $u_{\mathrm{tip}}=\dot{\Phi}$ in
the steady rotation regime since the wing length is normalized. We
choose the time scale such that the wingtip velocity is unity, thus
$T=1/\dot{\Phi}$. The first full rotation would thus be completed
at $t=6.68\,\left[T\right]$, but our computations are stopped at
$t=6$ to avoid the wing interacting with its own wake. As $t$ and
$\phi$ are equivalent, we use $\phi$ as it is more intuitive in
this case. The Reynolds numbers based on the terminal velocity of
the wing are the same as stated previously for the complete bumblebee,
and the Rossby number is defined as $Ro=R_{2}/c_{m}$ \citep{lentink2008}.
In our computations, $Ro=1.87$. For comparison, we also perform a
viscous simulation in which we multiply the viscosity by ten. Fig.
\ref{fig:setup}a illustrates the setup. The wing revolves around
a hinge placed at the center of a domain of size $4\times4\times2$
wing lengths, which is discretized using $1024\times1024\times512$
grid points. 

\subsubsection{Two flapping wings: the complete model\label{subsec:Two-flapping-wings}}

The previously described revolving setup is a simplification, and
differs in several aspects from an actual bumblebee. As the wing kinematics
in flapping flight is a periodic back-and-forth motion, each wingbeat
consists of two half strokes, usually termed up- and downstroke. In
each half stroke, a new leading edge vortex is created and shed as
a vortex puff at the stroke reversal. Describing the precise wingbeat
kinematics of insects is beyond the scope of this article, as it depends
on species, flight situation and varies between individuals. The wingbeat
motion is essentially parallel to the ground in hovering flight, while
this stroke plane is more inclined in forward flight. In hovering
flight, up- and downstroke are more symmetric than they are in forward
flight, which is the case we consider here. The incoming mean flow,
which in our simulations accounts for the insects forward flight velocity
(Galilean change of reference frame) acts differently on the wing
during the half-strokes. Nonetheless, in each half stroke, a leading
edge vortex is visible, as will be discussed later.

Our simulations take place in a $6R\times4R\times4R$ large, numerical
wind tunnel, which we resolve with $1152\times768\times768$ grid
points. The insect is tethered \textendash{} the imposed mean flow
accounts for its velocity; its wingbeat kinematics are prescribed.
Contrarily to the revolving wing case, we now use the wingbeat duration
to normalize time, $T=1/f$, as this is more natural in the flapping
configuration.

To model atmospheric turbulence, we use precomputed homogeneous isotropic
turbulence (HIT) as turbulent inflow. The resulting turbulent velocity
fluctuations can be added to the laminar inflow in a layer upstream
of the insect model. HIT is characterized by the turbulent kinetic
energy, the integral length scale and its Reynolds number. We vary
the turbulence intensity, $Tu=u_{\mathrm{RMS}}'/u_{\infty}$ defined
as the root mean square of velocity fluctuations normalized to flight
velocity, by altering the energy content of the turbulent perturbations
superimposed to the mean flow. The entire procedure allows us to study
insect flight from laminar to fully-developed turbulent flow regimes.
More details on this approach can be found in \citep{EKSLS15}.

\subsection{Numerical method\label{subsec:Numerical-method}}

Numerical simulations of the flow generated by insects have to face
two major challenges. First, as insects fly by flapping their wings,
the geometry of the problem is complicated and varies in time, implying
that the no-slip boundary condition for the Navier\textendash Stokes
equation has to be imposed on a complex fluid\textendash solid interface.
Second, many insects can be typically characterized by Reynolds numbers
in the intermediate regime \citep{Lissaman1983,Elli99}, i.e., $Re=\mathcal{O}\left(10^{3}\right)$.
In this Reynolds number regime, common simplifications, such as the
Stokes or inviscid approximations, are essentially nullified, leaving
us with the full non-linear unsteady problem. To cope with these challenges,
our numerical method combines the volume penalization method \citep{Angot1999}
with a Fourier pseudospectral discretization \citep{Schneider2005,Kolomenskiy2009},
for which we developed an open\textendash source computational environment,
available on Github\footnote{https://github.com/pseudospectators/FLUSI}
\citep{EKSS15}. The code solves the incompressible, penalized Navier\textendash Stokes
equations 
\begin{eqnarray}
\partial_{t}{\bm{u}}+{\bm{\omega}}\times{\bm{u}} & = & -\nabla\Pi+\nu\nabla^{2}{\bm{u}}-\underbrace{\frac{\chi}{C_{\eta}}\left({\bm{u}}-{\bm{u}}_{s}\right)}_{\text{penalization}}-\underbrace{\frac{1}{C_{\mathrm{sp}}}\nabla\times\frac{\left(\chi_{\mathrm{sp}}{\bm{\omega}}\right)}{\nabla^{2}}}_{\text{sponge}}\label{eq:PNST_org_momentum}\\
\nabla\cdot{\bm{u}} & = & 0\label{eq:PNST_divergence_free}\\
{\bm{u}}\left({\bm{x}},t=0\right) & = & {\bm{u}}_{0}\left({\bm{x}}\right)\qquad{\bm{x}}\in\Omega,t>0,\label{eq:PNST_inicond}
\end{eqnarray}
where the mask function $\chi\left({\bm{x}},t\right)$ is unity inside
the insect and zero otherwise and $C_{\eta}$ is the penalization
constant. The last term is a vorticity damping term used to gradually
damp vortices and alleviate the periodicity inherent to the Fourier
discretization. The role of this sponge is to relaminarize the (upwind)
flow as does the honeycomb in a windtunnel. Details on how the $\chi$
function and the solid body velocity field ${\bm{u}}_{s}$ are constructed
can be found in \citep{EKSS15}, along with a precise description
of the insect model and detailed validation tests. We use our code
only to compute DNS without additional turbulence modeling, and verify
via grid convergence studies that all spatial and temporal scales
are resolved. The penalization parameter is set to $C_{\eta}=5.66\cdot10^{-4}$,
and to determine $C_{\eta}$ as a function of the other parameters
in the simulations, the relation $K_{\eta}=\sqrt{\nu C_{\eta}}/\Delta x=0.074$
is used \citep{EKSS15}. This value of $K_{\eta}$ is used in all
reported simulations. A sponge layer with a thickness of $32$ grid
points and a damping constant $C_{\mathrm{sp}}=0.1$ is used to damp
the vorticity at the borders of the domain. For comparison, a second
simulation is performed increasing the viscosity by a factor of ten,
while keeping all other parameters constant. The accompanying paper
\citep{EKSS15} contains more details on the method, as well as a
large variety of validation tests.

\subsection{Helicity, helicity spectra and scale-dependent helicity\label{subsec:Helicity,-helicity-spectra}}

Helicity is a quantity introduced by Betchov in \citep{Betchov1961},
important to study the dynamics of turbulent flows. In \citep{More61,Moff69}
it was shown that energy and helicity are two conserved quantities
of the incompressible Euler equations. For a comprehensive review
on helicity we refer to \citep{MoTs92}. Considering the velocity
field ${\bm{u}}$ and the corresponding vorticity ${\bm{\omega}}=\nabla\times{\bm{u}}$,
the kinetic helicity, $H({\bm{x}})={\bm{u}}\cdot{\bm{\omega}}$, can
be defined, see, e.g., \citep{More61,Moff69}. The helicity yields
a measure of the geometrical statistics of a turbulent flow and allows
us to quantify its chirality. It changes sign when applying a mirror
symmetry to the reference frame (transforming it from left to right
handed). Integrating the helicity over space and dividing it by the
volume one obtains the mean helicity $\left\langle H\right\rangle =\langle{\bm{u}}\cdot{\bm{\omega}}\rangle$.

The relative helicity 
\begin{equation}
h({\bm{x}})=\frac{H}{|{\bm{u}}|\,|{\bm{\omega}}|}
\end{equation}
corresponds to the cosine of the angle between the velocity and the
vorticity at each spatial position. The range of $h$ thus lies between
$-1$ and $+1$, corresponding to anti-alignment and alignment of
the velocity and the vorticity vector, respectively. \medskip{}
\begin{center}
\emph{Energy and helicity balance equation}
\par\end{center}

\medskip{}

Similar to the dissipation of energy (in the absence of forcing),
$d_{t}\left\langle E\right\rangle =-2\nu\left\langle Z\right\rangle $
where $\left\langle E\right\rangle =\left\langle |{\bm{u}}|^{2}\right\rangle /2$
and $\left\langle Z\right\rangle =\left\langle |{\bm{\omega}}|^{2}\right\rangle /2$
are respectively the mean energy and enstrophy, mean helicity satisfies
a balance equation, 
\begin{equation}
d_{t}\left\langle H\right\rangle \;=\;-2\nu\left\langle H_{\omega}\right\rangle 
\end{equation}
where $\left\langle H_{\omega}\right\rangle =\langle{\bm{\omega}}\cdot({\nabla\times\bm{\omega}})\rangle$
is the mean helicity of vorticity (also called superhelicity) assuming
absence of helical forcing. In viscous flows, helicity is generated
and dissipated, while in the inviscid case ($\nu=0$) the Euler equations
conserve the mean kinetic helicity. Contrary to energy neither helicity
of velocity nor helicity of vorticity are positive definite quantities.
The point-wise helicity $H({\bm{x}},t)$ of velocity satisfies the
equation \citep{KTMa04}, 
\begin{equation}
\partial_{t}H+{\bm{u}}\cdot\nabla H\;=\;-\nabla\cdot({\bm{\omega}}p)+\frac{1}{2}\nabla\cdot({\bm{\omega}}|{\bm{u}}|^{2})+\nu(\nabla^{2}H-2(\nabla{\bm{u}}\nabla{\bm{\omega}}))
\end{equation}
This shows that for the helicity dynamics both the nonlinear and the
viscous terms locally play a role, either in enhancing or diminishing
the helicity.

\medskip{}
\begin{center}
\emph{Energy and helicity spectrum}
\par\end{center}

\medskip{}

Computing the Fourier transform of the velocity and the vorticity,
denoted by $\widehat{\cdot}$, the isotropic energy and helicity spectra
can be defined, 
\begin{equation}
E(k)\,=\,\frac{1}{2}\,\sum_{k=|{\bm{k}}|}|\widehat{\bm{u}}({\bm{k}})|^{2}\quad,\quad H({k})\,=\,\,\sum_{k=|{\bm{k}}|}\widehat{\bm{u}}({\bm{k}})\cdot\widehat{\bm{\omega}}(-{\bm{k}})\;.
\end{equation}
Note that $H(k)$ is also real valued, but a signed quantity, and
by construction we have $\sum_{k\ge0}E(k)=E$ and $\sum_{k\ge0}H(k)=\left\langle H\right\rangle $
which justifies that $E(k)$ and $H(k)$ are called the spectral density
of energy and helicity, respectively. Applying the Cauchy-Schwarz
inequality, it follows that $|H(k)|\le2kE(k)$, which motivates the
introduction of the relative helicity spectrum $|H(k)|/(2kE(k))\le1$.
In \citep{KTMa04} it has been shown to fall off linearly in wave-number
for large $k$, restoring thus the mirror symmetry of the flow at
small scales in the case of isotropic turbulence.

\medskip{}
\begin{center}
\emph{Scale-dependent energy and helicity}
\par\end{center}

\medskip{}

The vorticity and velocity field can be decomposed into an orthogonal
wavelet series, i.e. for the velocity we have
\[
\bm{u}=\sum_{\mu}\sum_{\bm{i}}\sum_{j}\widetilde{\bm{u}}_{\mu,\bm{i},j}\psi_{\mu,\bm{i},j}\left(\bm{x}\right)
\]
where $j$ is the scale index, $\mu$ the direction index and $\bm{i}$
is the position vector. The coefficients $\widetilde{\bm{u}}_{\mu,\bm{i},j}=\left\langle \bm{u},\psi\right\rangle $
are then called the wavelet transform of $\bm{u}$, where $\psi$
is the wavelet. Orthogonal wavelets typically do not posses a closed-from
expression, but they are rather defined in terms of quadrature-mirror
filters. The contributions at scale $j$ can be obtained (for details
see, e.g., \citep{FaSc15}) by summing over all scales and directions
for a given scale $j$:
\[
\bm{u}_{j}(\bm{x})=\sum_{\bm{i}}\sum_{\mu}\widetilde{\bm{u}}_{\mu,\bm{i},j}\psi_{\mu,\bm{i},j}\left(\bm{x}\right)
\]
which corresponds essentially to bandpass filtering since all other
scales are set to zero. For the vorticity the above decomposition
can be applied analogously. The scale-dependent energy can thus be
defined as 
\begin{equation}
E_{j}({\bm{x}})\,=\,\frac{1}{2}\,{\bm{u}}_{j}({\bm{x}})\cdot{\bm{u}}_{j}({\bm{x}})
\end{equation}
and integrating over ${\bm{x}}$ yields the mean energy $\left\langle E_{j}\right\rangle $
at scale $2^{-j}$, which is called energy scalogram. Summing $\left\langle E_{j}\right\rangle $
over scale we obtain the total energy $E=\sum_{j}\,\langle E_{j}\rangle$.

Analoguously the scale-dependent helicity can be defined as 
\begin{equation}
H_{j}({\bm{x}})\,=\,{\bm{u}}_{j}({\bm{x}})\cdot{\bm{\omega}}_{j}({\bm{x}})
\end{equation}
which was introduced in \citep{YOSKF09} in the context of isotropic
turbulence. 
The scale-dependent helicity preserves Galilean invariance, though
the kinetic helicity itself does not. Integrating $H_{j}$ over ${\bm{x}}$
yields the mean helicity $\left\langle H_{j}\right\rangle $ at scale
$2^{-j}$, which we call helicity scalogram. The corresponding mean
helicity is obtained by summing $\left\langle H_{j}\right\rangle $
over scale, $\overline{H}=\sum_{j}\,\langle H_{j}\rangle$, due to
the orthogonality of the wavelet decomposition.

The scale-dependent relative helicity can be defined correspondingly
as 
\begin{equation}
h_{j}({\bm{x}})=\frac{H_{j}}{|{\bm{u}}_{j}|\,|{\bm{\omega}}_{j}|}
\end{equation}
and can be used to analyze the probability distribution of the cosine
of the alignment angle \citep{YOSKF09}. 

The scale $2^{-j}$ can be related to the wavenumber $k_{j}$ as 
\begin{equation}
k_{j}=k_{\psi}2^{j},\label{coifkj}
\end{equation}
where $k_{\psi}=\int_{0}^{\infty}k|\hat{\psi}(k)|dk/\int_{0}^{\infty}|\widehat{\psi}(k)|dk$
is the centroid wavenumber of the chosen wavelet ($k_{\psi}=0.77$
for the Coiflet 12 used here). Thus the scale-dependent energy and
helicity can be directly related to their corresponding Fourier spectra.

The wavelet energy spectrum can be obtained using the scalogram and
eq. (\ref{coifkj}), 
\begin{equation}
{\widetilde{E}}(k_{j})=\frac{1}{2\Delta k_{j}}\langle E_{j}\rangle,\label{wave_spe}
\end{equation}
where $\Delta k_{j}=(k_{j+1}-k_{j})\ln2$ \citep{Mene91,Addison}.
It is thus directly related to the Fourier energy spectrum and yields
a smoothed version~\citep{Farge92,Mene91}. The orthogonality of
the wavelets with respect to scale and direction guarantees that the
total energy is obtained by direct summation, $E=\sum_{j}{\widetilde{E}}(k_{j})$.

The wavelet helicity spectrum can then be obtained likewise 
\begin{equation}
{\widetilde{H}}(k_{j})=\frac{1}{2\Delta k_{j}}\langle H_{j}\rangle,\label{wave_hel_spe}
\end{equation}
and again summation over $j$ yields the total mean helicity. We anticipate
that the wavelet helicity spectrum is a smoothed version of the Fourier
helicity spectrum.



The spatial variability of the wavelet energy and helicity spectra
at a given wavenumber $k_{j}$ can be quantified by the standard deviation,
defined as 
\begin{eqnarray}
{\sigma}[E_{j}]=\frac{1}{2\Delta k_{j}}\sqrt{\langle({\bm{u}}_{j}\cdot{\bm{u}}_{j})^{2}\rangle-\left(E_{j}\right)^{2}}\quad,\quad{\sigma}[H_{j}]=\frac{1}{2\Delta k_{j}}\sqrt{\langle({\bm{u}}_{j}\cdot{\bm{\omega}}_{j})^{2}\rangle-\left(H_{j}\right)^{2}}\,.\label{wl_variace3d}
\end{eqnarray}
Thus the flow intermittency can be quantified. This is not possible
using Fourier spectra as all spatial information is lost. The spatial
variability of the energy spectrum can be related to the scale-dependent
flatness, defined as the ratio of the fourth- to the second-order
moment of the scale dependent velocity, as discussed, e.g., in~\citep{FaSc15}.
Increasing flatness values for decreasing scale, i.e., values larger
than three which are obtained for a Gaussian distribution, are attributed
to the flow intermittency. 

\begin{figure}
\begin{centering}
\includegraphics[width=0.9\textwidth]{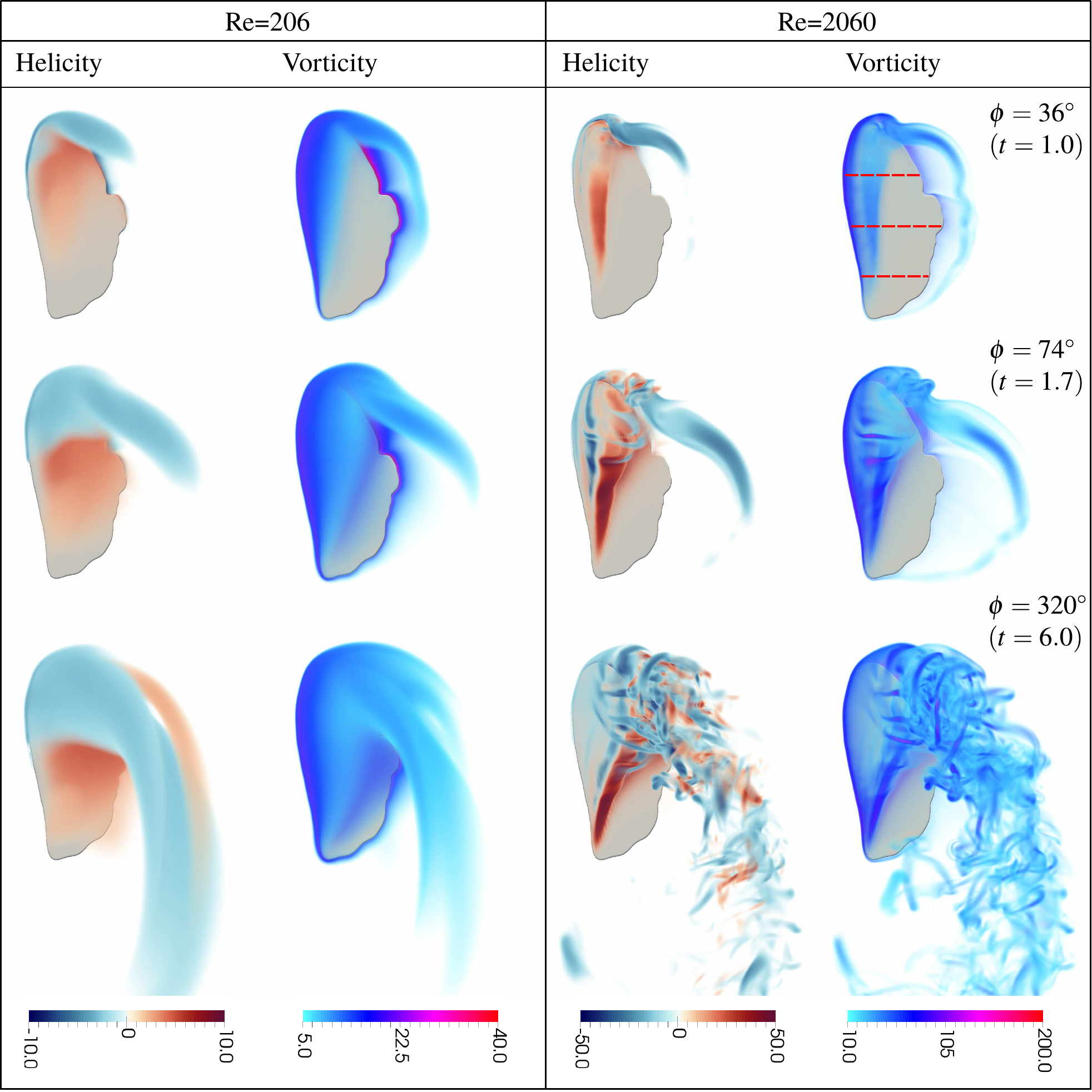}
\par\end{centering}
\caption{Flows generated by a rotating wing at two different Reynolds numbers,
$Re=206$ (left) and $2060$ (right), visualized by their helicity
$H(\bm{x})$ and vorticity magnitude $|\bm{\omega}(\bm{x})|$ at three
rotation angles $\phi$ (rows). The view is rotated such that the
observer looks in the direction of the wing normal. The flow topology
becomes more complex when the Reynolds number increases. Essential
features, such as leading edge and tip vortices, are observed in both
cases, but remain stable at $Re=206$ and develop strong instabilities
at $Re=2060$. All quantities are dimensionless, i.e. helicity is
given in $\left[L/T^{2}\right]$ and vorticity magnitude in $\left[1/T\right]$.
Red dashed lines correspond to positions of 2D slices shown in Fig.
\ref{fig:2dcuts}.}
\label{fig:oldfig2}
\end{figure}

\section{Numerical results\label{sec:Numerical-results}}

The results in this section are all presented in dimensionless form,
using the winglength $R$, the mass scale $\varrho_{f}R^{3}$ (therefore
the dimensionless fluid density is unity) and a time scale $T$. The
latter depends on the context: in the revolving wing, we set $u_{tip}$
to $1$ and thus $T=R/u_{tip}$, while in the bumblebee setup the
cycle duration is a more convenient parameter for normalization.

\subsection{Flow generated by a revolving bumblebee wing\label{subsec:results-revolving}}

This section deals with a flow generated by a bumblebee wing which
steadily revolves around a hinge point with a constant angle of attack,
see Fig.~\ref{fig:setup} (a). The setup is inspired by experimental
contributions considering revolving wings of either rectangular \citep{Jones2016}
or insect-inspired shape \citep{Pick2009}. In particular, \citep{Jones2016}
focuses on the bursting of the leading edge vortex, i.e. its transition
from simple to complex topology. The authors find that the bursting
does not have an impact on the aerodynamic force. The vortex can burst
but still remains attached to the wing, i.e., there is no LEV separation,
and helicity can be used to characterize this bursting. 

We first visualize, in Fig \ref{fig:oldfig2}, the helicity $H(\bm{x})$
and vorticity magnitude $|\bm{\omega}(\bm{x})|$, at three different
instants, for two different flows, corresponding to $Re=206$ and
$2060$, respectively. For both flows the wing motion starts from
rest in a quiescent fluid and a vortex is formed. 

The first time instant, $\phi=36^{\circ}$ (top row), corresponds
to the early phase of steady rotation. In the left part of Fig. \ref{fig:oldfig2},
the simulation with $Re=206$ is visualized and the right part corresponds
to the $Re=2060$ case, which is the Reynolds number of the bumblebee.
In both simulations, large amounts of vorticity are created at the
leading edge, where the flow separates due to the elevated angle of
attack. Thus, in both cases, an LEV is formed, but the quantitative
scale for vorticity is reduced in the viscous case. A tip vortex forms
as well in both cases.

At the same time, the visualization of kinetic helicity show that
virtually no helicity is generated at the leading edge, even though
large amounts of vorticity are available. This indicates the lack
of spanwise flow at the leading edge. The axial flow seems to develop
further away form the leading edge, which is also where Fig. \ref{fig:oldfig2}
shows positive helicity, again at different magnitudes for the two
Reynolds numbers. In the higher Reynolds number case, the region of
positive helicity is more strongly confined and marks a distinct vortex
core. In both cases, the wing tip vortex features negative helicity.
The topological reconnection of the LEV and the tip vortex contains
a curious transition from the positive helicity in the LEV core to
the overall negative helicity in the wing tip vortex. 

The outwards axial flow in the LEV is driven by the centrifugal force
and the axial pressure gradient produced the conical shape of the
vortex. The axial flow in the wing tip vortex is created by entrainment
of the fluid behind the moving wing. Consequently, the helicity changes
sign near the wing tip. 

Later on, at $\phi=320^{\circ}$ (bottom row), the differences in
the vorticity fields of the two cases become quite remarkable, as
the higher Reynolds number case develops much finer flow features
near the wing tip, which are inhibited by the viscosity in the other
case. It is also noted that a coherent leading edge vortex is visually
less easily defined in the low viscosity case. The visualization of
kinetic helicity $H\left({\bm{x}}\right)={\bm{u}}\cdot{\bm{\omega}}$
in Fig.~\ref{fig:oldfig2} looks qualitatively similar to the vorticity
magnitude regarding the appearance of fine structures. The tip vortex
is helical with a negative value of $H$, while the region near the
root until midspan features positive values of $H$. In the high Reynolds
case, a strongly helical leading edge vortex is visible at $\phi=75^{\circ}$,
which becomes incoherent towards the tip. At $\phi=320^{\circ}$,
more than half of the wing features an incoherent, burst leading edge
vortex. We note at either Reynolds number that no vortex shedding
occurs, meaning that the leading edge vortex remains attached to the
wing.

\begin{figure}
\begin{centering}
\includegraphics[width=0.8\textwidth]{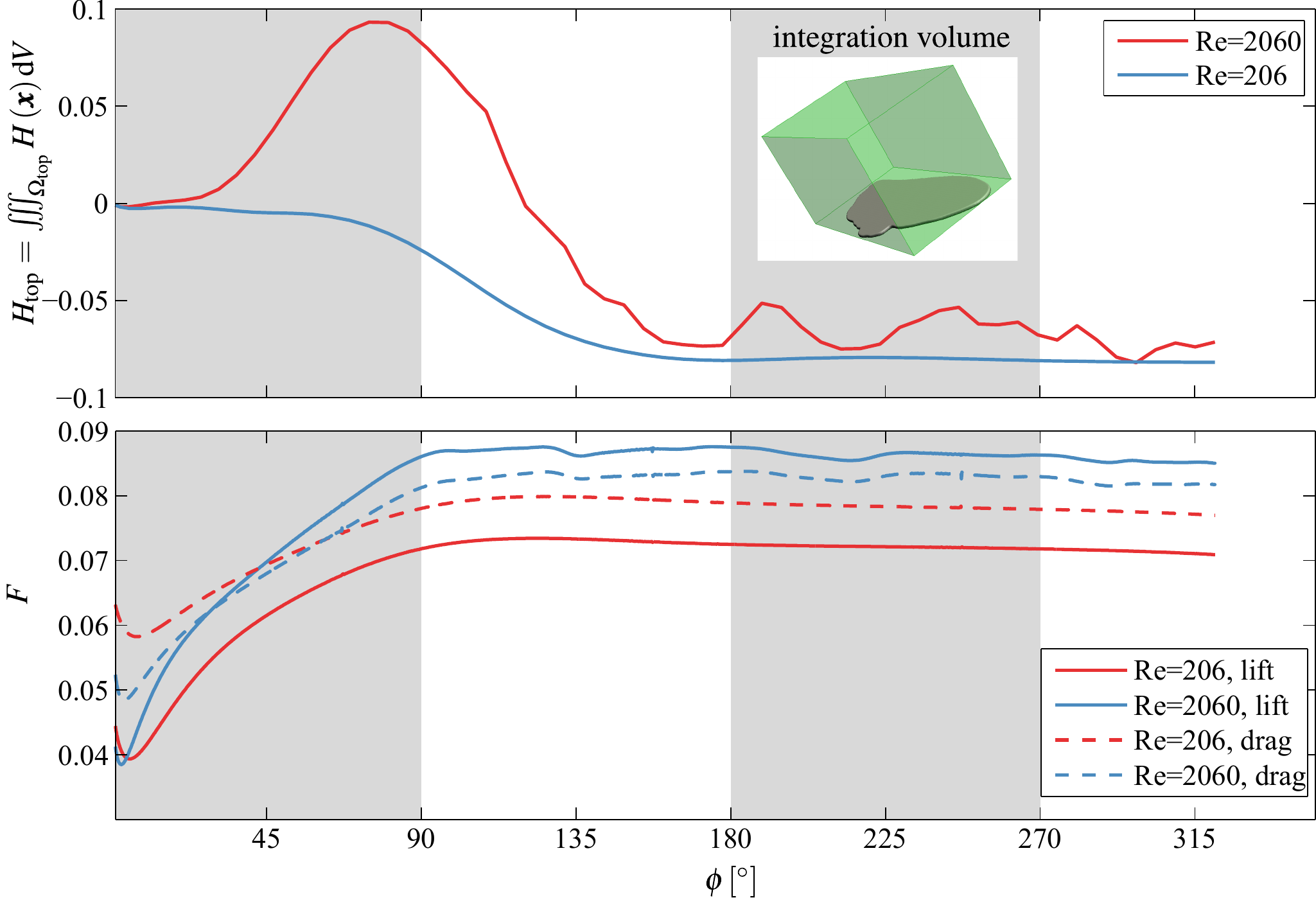}
\par\end{centering}
\caption{Evolution of helicity, lift and drag as a function of the rotation
angle $\phi$ for the revolving wing at $Re=206$ and $2060$. Top
part shows integral helicity $H_{\mathrm{top}}=\iiint_{\Omega_{top}}H\left(\bm{x}\right)\,\mathrm{d}V$,
where $\Omega_{\mathrm{top}}=\left[-0.35,+0.15\right]\times\left[0,1\right]\times\left[0,1\right]$
is a cubic control volume on the top surface (suction side) of the
wing, as shown in the inset. Bottom part shows the aerodynamic force,
split into lift (in the $z$-direction of the laboratory system) and
drag (the magnitude of the $x$- and $y$ component). All quantities
are dimensionless, i.e. $F$ is given in $\left[ML/T^{2}\right]$
and $H_{\mathrm{top}}$ in $\left[L^{4}/T^{2}\right]$.\label{fig:time-evolution-rotwing}}
\end{figure}

This LEV bursting becomes more clearly visible when integrating the
helicity density over a control volume above the suction side of the
wing, where the leading edge vortex is found. This value is shown
in Fig. \ref{fig:time-evolution-rotwing} (top). From vanishing helicity
due to the quiescent initial condition, the integral value $H_{\mathrm{top}}=\iiint_{\Omega_{\mathrm{top}}}H\left(\bm{x}\right)\,\mathrm{d}V$
follows a qualitatively different evolution for the two Reynolds numbers
considered. In the viscous case, $H_{\mathrm{top}}$ is negative throughout
the simulation and builds up until around $\phi=180^{\circ}$, remaining
constant around $-0.08$ afterwards. By contrast, the high Reynolds
number flow first builds up positive $H$ until a maximum is reached
at $\phi=81^{\circ}$, then rapidly drops to a constant, negative
value very close to the viscous case. The breakdown of positive helicity
is a consequence of vortex bursting.

As emphasized in \citep{Jones2016}, the consequences for the force
production are marginal. Fig.~\ref{fig:time-evolution-rotwing} (bottom)
shows the lift and drag component of the aerodynamic force, which
is computed as $\bm{F}=\int\chi\left(\bm{u}-\bm{u}_{s}\right)/C_{\eta}\,\mathrm{d}V$
\citep{Angot1999,EKSS15}. Their evolution with the rotation angle
is qualitatively similar, and an almost steady force is produced after
$\phi=90^{\circ}$, with only small fluctuations in the higher Reynolds
number case. The qualitatively very different behavior in $H_{\mathrm{top}}$
is thus not reflected in the force production. Interestingly, though
we varied the Reynolds number by a factor of ten, the lift force changes
only by $16\,\%$, but the viscous case produces quantitatively more
drag than lift, which is the opposite of the higher Reynolds number
case.

\begin{figure}
\begin{centering}
\includegraphics[width=0.8\textwidth]{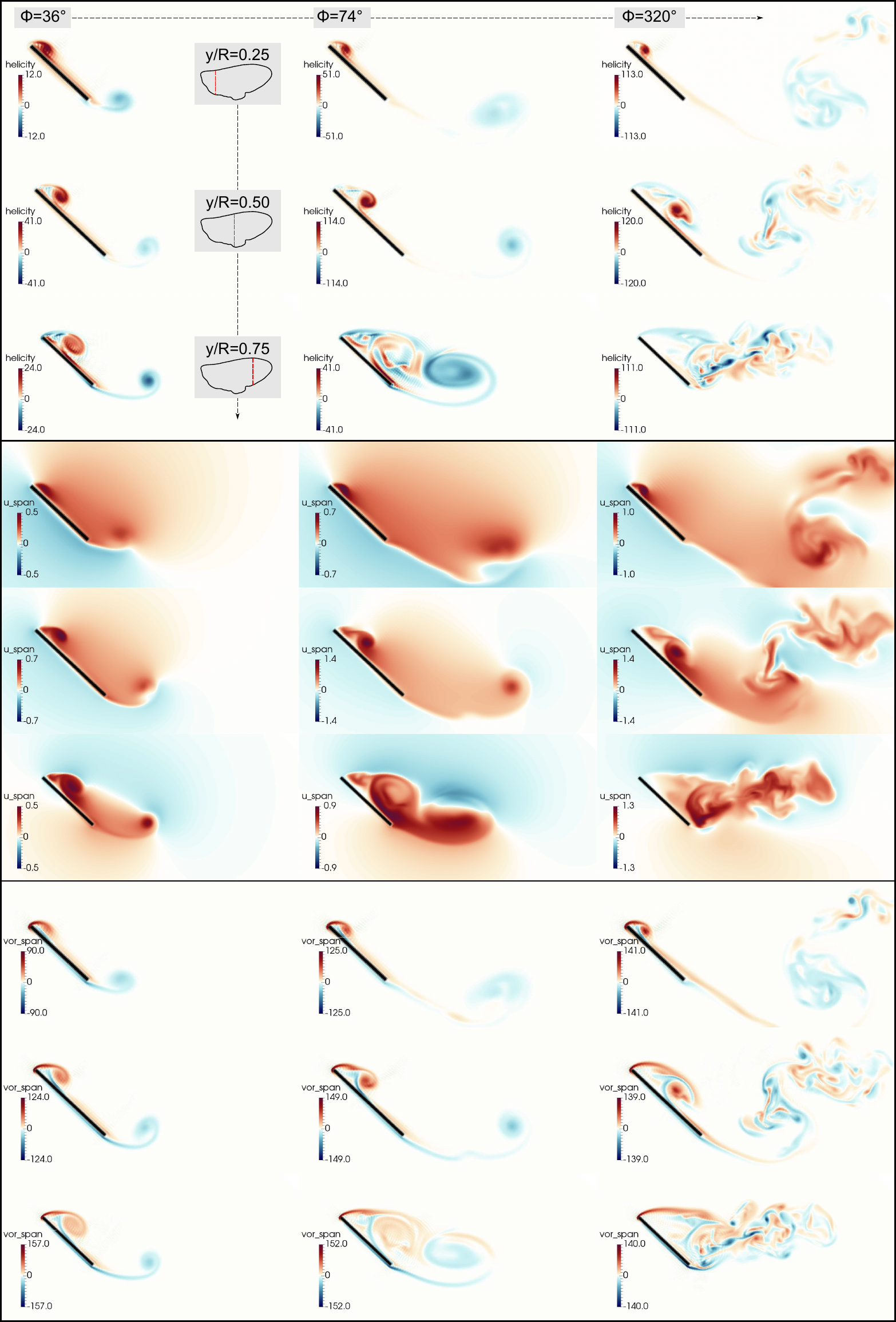}
\par\end{centering}
\caption{Flow around a rotating wing, $Re=2060$. Figure shows two-dimensional
slices of helicity (top box), spanwise velocity (middle box) and spanwise
vorticity (bottom box). Slices are at three different spanwise positions,
$y/R=0.25,0.50,0.75$ (rows, visualized by insets in top part) and
rotation angles $\phi=36^{\circ},\,74^{\circ},\,320^{\circ}$ (columns).
All quantities are dimensionless, i.e. helicity is given in $\left[L/T^{2}\right]$,
velocity in $\left[L/T\right]$ and vorticity in $\left[1/T\right]$.\label{fig:2dcuts}}
\end{figure}

Figure~\ref{fig:2dcuts} displays two-dimensional sectional plots
of kinetic helicity, axial flow and axial vorticity at three time
rotation angles for three different spanwise positions. At $\phi=36^{\circ}$,
we observe the formation of a conical LEV core above the suction side
of the wing at all three spanwise locations. Large positive spanwise
vorticity in the core is collocated with large outwards spanwise velocity,
yielding large positive helicity density. In the LEV feeding sheet,
however, the helicity is already changing sign from positive over
the proximal part to negative over the distal part of the wing. This
may be an early sign of the developing breakdown instability. At $\phi=74^{\circ}$,
the proximal part of the wing still supports a compact conical core.
However, the LEV core bursts over the distal part. This is seen by
thickening of the core and and emergence of smaller secondary structures
that wrap around the primary core. The helicity is still positive,
but not as large as before the burst. 

At the final rotation angle, $\phi=320^{\circ}$, the LEV is in its
statistical equilibrium state. It begins as a laminar conical vortex
from the root of the wing and bursts at around 2/3 of the wing length,
forming a series of 3d strongly helical trailing vortices (ribs) which
are perpendicular to the LEV.

Note that our results are essentially not frame dependent, because
the vorticity associated with changing between the laboratory reference
frame and a moving reference frame of the wing is of order 1, but
the vorticity in the vortices is of order 100, i.e., two orders of
magnitude larger.

Garmann and Visbal~\citep{Garmann2014} point out the co-existence
of the burst instability of the LEV core and the Kelvin\textendash Helmholtz
instability in the feeding LEV sheet. While the LEV burst is obvious
in our numerical simulations, the Kelvin\textendash Helmholtz instability
is not apparent, possibly because the shear layer transition point
is too far from the rotation axis at $Re=2060$. The two instabilities
may have different scaling with the Reynolds number, and this question
needs further investigation. 

\begin{figure}
\begin{centering}
\includegraphics[width=1\textwidth]{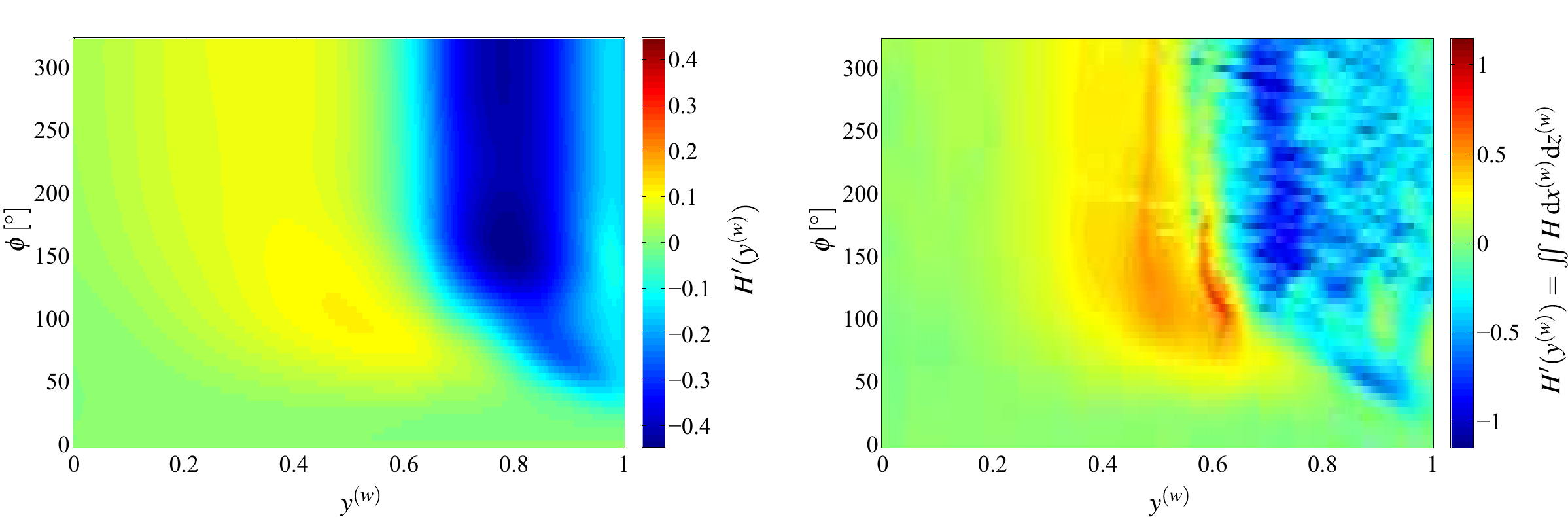}
\par\end{centering}
\caption{Helicity $H'=\iint_{\Omega}H\,\mathrm{d}x^{(w)}\mathrm{d}z^{(w)}$,
which is the integral of helicity $H\left(\bm{x}\right)$ in slices
normal to the wing in chordwise direction ($y^{(w)}=\mathrm{const}$,
see Fig. \ref{fig:setup}b for axis definition), as a function of
the distance to wing root $y^{(w)}$ and rotation angle $\phi$. The
integration domain is $\Omega=[-0.35,\,0.15]\times[0,1]$. All quantities
are dimensionless, i.e. $y^{(w)}$ is given in $[L]$.\label{fig:axialflow_sketch}}
\end{figure}

\begin{figure}
\begin{centering}
\includegraphics[width=1\textwidth]{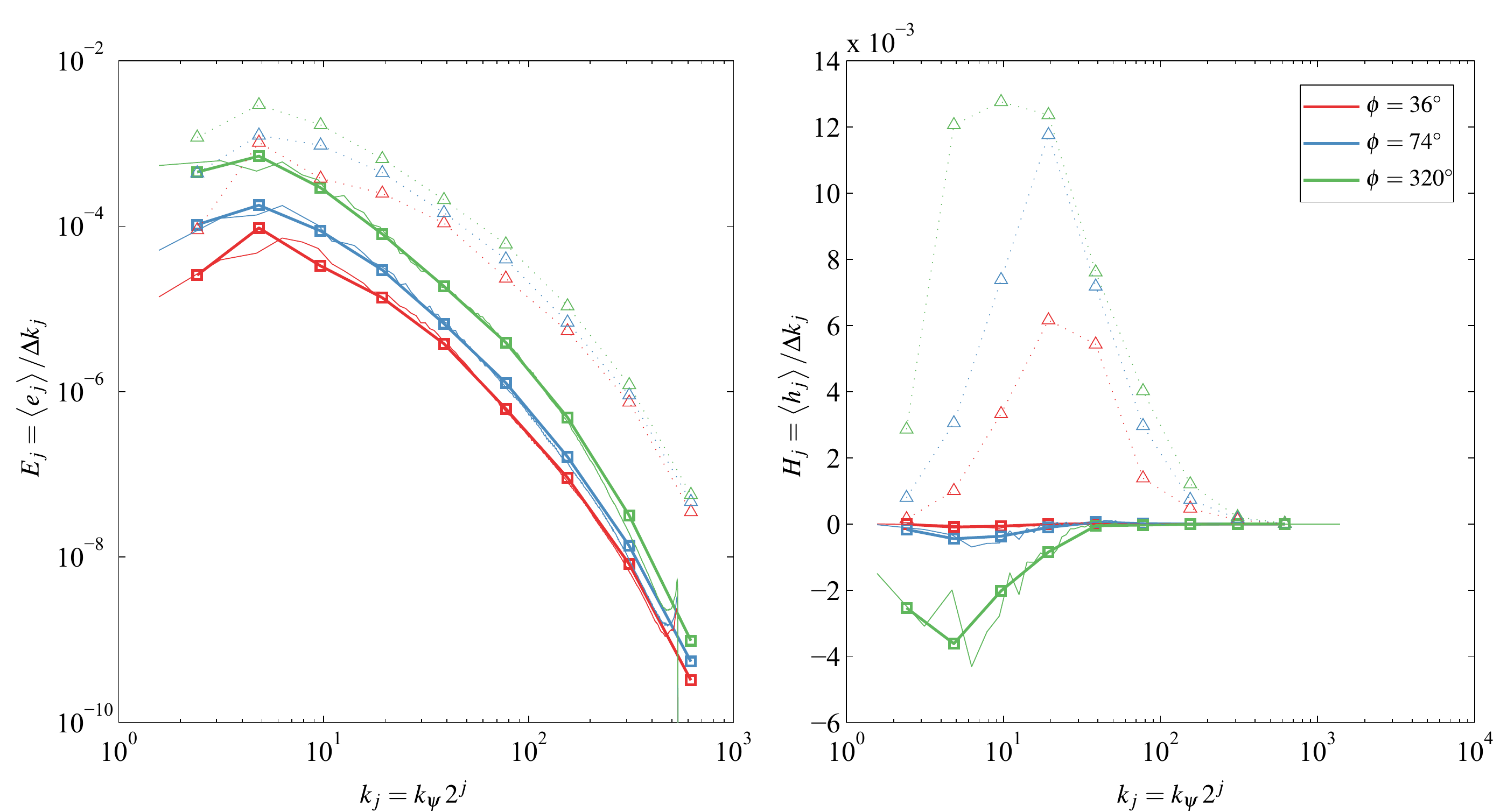}
\par\end{centering}
\caption{Scalograms of energy and helicity in the rotating wing at $Re=2060$:
Wavelet energy (left) and helicity (right) spectra (thick, continuous
lines) together with their corresponding standard deviation (dashed
lines) at three different rotation angles, $\phi=36^{\circ},\,74^{\circ}$
and $320^{\circ}$, computed using orthogonal Coiflet 12 wavelets.
For comparison, the Fourier spectra are also shown (thin, continuous
lines).\label{fig:helspectra}}
\end{figure}

Figure~\ref{fig:axialflow_sketch} illustrates the evolution of relative
helicity as a function of spanwise position for $Re=206$ (left) and
$Re=2060$ (right). The horizontal axis in each of the panels corresponds
to the spanwise position $y^{(w)}$ (see Fig. \ref{fig:setup}b for
the axis definition), and the vertical axis corresponds to the rotation
angle $\phi$. Thus, the color of a selected row of pixels on the
diagram shows how the helicity density varies along the wing at a
given $\phi$. A column of pixels, by contrast, shows how the helicity
density at a given $y^{(w)}$ varies in time. We first discuss the
low $Re$ case. At startup, $\phi<25^{\circ}$, the helicity density
is negligibly small, which means that, even though some strong vorticity
may be produced at the sharp edges, no significant axial flow has
developed in the vortex cores. After $\phi=25^{\circ}$, the positive
helicity builds up in the LEV, and negative helicity builds up in
the wing tip vortex. The wing tip vortex expands as time progresses,
until saturation after $\phi=150^{\circ}$.

At the larger $Re$, the diagram is similar to the extent that helicity
is positive in the LEV, negative in the wing tip vortex, and the two
regions develop in time until saturation at about the same time $\phi=150^{\circ}$
and the same radial position $y^{(w)}=0.55$. However, the magnitude
of helicity is about 3 times as large compared to the low-$Re$ case.
This is probably related to the enhanced axial flow in the high-$Re$
LEV, and overall larger vorticity production in that case. 

The wavelet energy spectra (Fig.~\ref{fig:helspectra}, left) in
log-log representation and helicity spectra in lin-log representation
(Fig.~\ref{fig:helspectra}, right) show the scale distribution of
energy and helicity, respectively. They yield similar information
as the Fourier spectra, however the wavelet spectra are less influenced
by the mask function $\chi\left(\bm{x},t\right)$, in particular at
small scales, used in the computations to impose the no-slip boundary
conditions. We observe that both energy and helicity values grow in
time and that the maximum magnitude is at the same wavenumber, $k=5$,
where also a peak in the kinetic energy is observed. The corresponding
standard deviations (dashed lines) illustrate the spatial fluctuations
of energy and helicity. We find that small energy and helicity values
at large $k$ exhibit nevertheless large fluctuations, which is a
signature of the flow intermittency. 

\medskip{}

\begin{figure}
\begin{centering}
\includegraphics[width=0.8\textwidth]{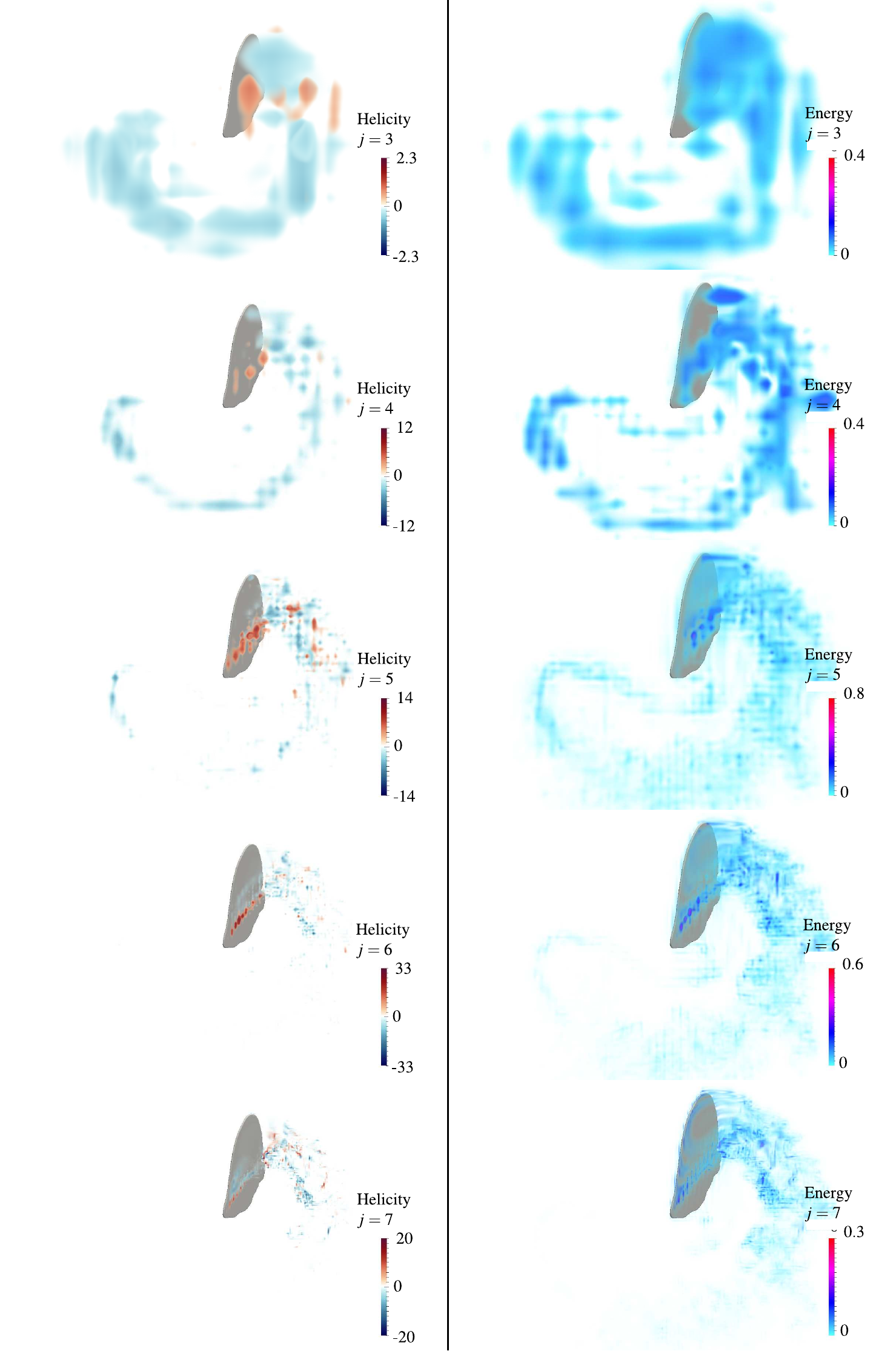}
\par\end{centering}
\caption{Rotating wing at $Re=2060$ and the terminal rotation angle $\phi=320^{\circ}$.
Helicity (left) and energy (right) are visualized at four different
scales from $j=3$ (large scale) to $j=7$ (small scale) (from top
to bottom). Coiflet 12 orthogonal wavelets were used for scale extraction.\textcolor{blue}{{}
}\label{fig:helscalewise}}
\end{figure}

Visualizations of the scale-wise helicity together with the energy
are presented in Figure~\ref{fig:helscalewise} at $t=6$. The (positively)
helical leading edge vortex is well visible at scales $2^{-5}$ and
$2^{-6}$, while the tip vortex, visible at larger scales, is predominantly
negative. However, positively helical structures are also present
in the tip vortex at all scales. Also, a negatively helical secondary
LEV core is visible, adjacent to the primary positive LEV at scales
$2^{-6}$ and $2^{-7}$. The secondary core is rotating in opposite
direction of the primary core, see, e.g., \citep{Garmann2014}. Fine
scaled energy contributions are located near the wing, while the far
field features energy at relatively larger scales. This is not surprising,
since vortical structures at smaller scales decay faster because of
viscous dissipation.

\cleardoublepage{}


\subsection{Flow generated by flapping wings of a tethered bumblebee}

From the revolving wing studied in the previous section we now proceed
to the case of a bumblebee. A key advantage of our numerical method
is the simplicity with which complex geometries can be taken into
account. Therefore we include the insect's body in the computational
model, including its legs, antennae and proboscis. For an illustration
we refer to Fig.~\ref{fig:setup} (c). The body is responsible for
the major part of aerodynamic drag and it may contribute, though less
significantly, to the lift as well. In the interest of brevity we
refer to the supplementary material of \citep{EKSLS15} for a complete
description of the modeled insects morphology. 

\begin{figure}
\centering{} 
\includegraphics[width=1\textwidth]{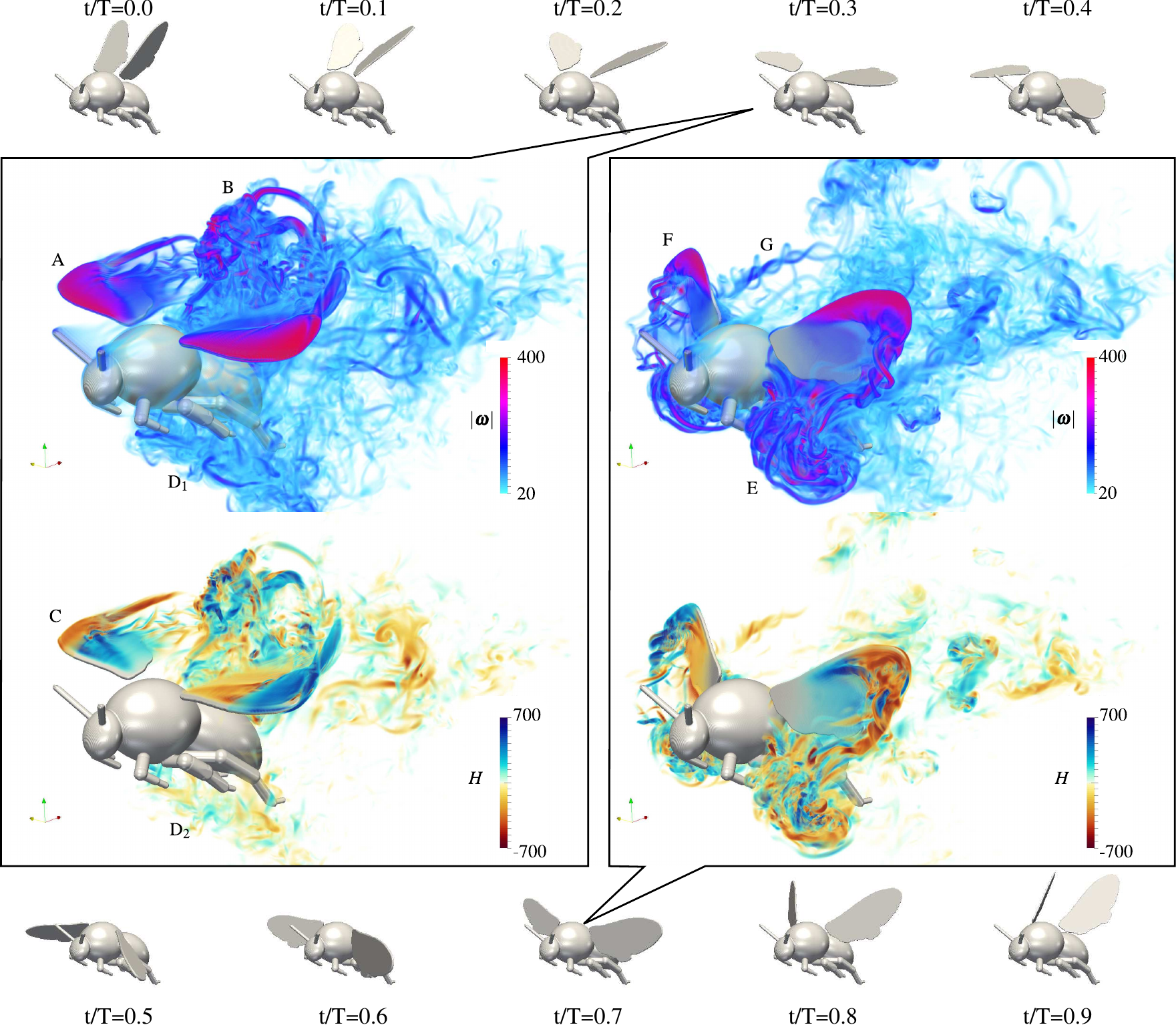}\caption{
Bumblebee in laminar inflow. Shown is the wingbeat kinematics for
the down- (top) and upstroke (bottom). For two selected times during
down- and upstroke, the flowfield is visualized by vorticity magnitude
(top) and helicity (bottom). Helical leading edge vortices can be
identified.}
 \label{fig_bb_helicity} 
\end{figure}

Fig. \ref{fig_bb_helicity} illustrates, in the top and bottom strips,
the wingbeat kinematics for the down- and upstroke. The mean stroke
plane is inclined with respect to horizontal, and the geometric angle
of attack is larger during the downstroke. 

Fig. \ref{fig_bb_helicity} also shows visualizations of the flow
field at two selected instants, $t=0.3$ and $t=0.7$ , which are
in the middle of the down- and upstroke, respectively. The vorticity
field, $\left|\bm{\omega}\right|$, shows the large amount of vorticity
generated at the wing's leading edges (A). This zone of intense vorticity
appears to be continuous even in the tip vortex. Behind the body,
where the wings shed their leading edge vortices at the end of the
previous upstroke (B), another zone with elevated values of vorticity
exists. The overall flow topology is highly complex, but symmetry
is not broken. The reason for this symmetry lies in the precision
of the numerical method, in which no symmetry-breaking perturbations
occur. The visualization of helicity however shows that leading edge
and tip vortex can be clearly distinguished (C) as they have opposite
signs in $H$. This distribution of vorticity and helicity is qualitatively
similar to what has been found in the case of the revolving wing in
Fig. \ref{fig:oldfig2}, though in the quantitative scales of $H$
and $\left|\bm{\omega}\right|$ differ significantly, even for a comparable
Reynolds number. However the Reynolds number is difficult to compare
in both cases, as the wingtip velocity in the bumblebee case is not
constant. Instead, the cycle-averaged value is used, but this implies
that the instantaneous wingtip velocity can be larger than in the
revolving case. In addition, the mean flow, which was not present
in the revolving wing, increases the instantaneous relative velocity
during the downstroke, as the wings move upstream.

Remarkably, many regions containing vorticity away from the insects
exhibit less helicity, cf. $\mathrm{D}_{1}$ and $\mathrm{D}_{2}$.
As discussed previously, this implies that the non-linearity of the
Navier\textendash Stokes equation is strong in those regions, and
that these structure participate more in the Kolmogorov cascade of
energy.

At the end of the downstroke, the wing reverses its direction, and
the leading edge vortex is shed into the wake. The resulting vortex
'puff' can be seen in the second visualized time instant (E). The
puff features a much more complex topology than the leading edge vortex
(A), and its helicity has no preferential sign. A new, though weaker
leading edge vortex is formed (F) at the wings, with the same pattern
of helicity and vorticity. The wingtip vortex from the downstroke
has formed two vortex filaments that form a helix (G).

We thus note that the LEV has similar features to the one produced
in a rotating wing, but the wake topology differs due to the reciprocal
flapping motion.

We now proceed and revisit the model bumblebee in turbulent inflow,
in the same manner as has been done in previous work \citep{EKSLS15}.
The inflow turbulence, imposed in a layer upstream of the insect,
consists of velocity fluctuations $\bm{u}'$ added to the mean inflow
$\bm{u}_{\infty}$. The fluctuations are obtained from pre-computed
simulations of homogeneous isotropic turbulence (HIT) with Reynolds
numbers $R_{\lambda}$ ranging 90 to 230. Scaled to the insect dimensions
this yields turbulence intensities $Tu=u'_{\mathrm{RMS}}/u_{\infty}$
between 0.17 and 0.99. For all turbulence intensities, a single realization
is, due to the erratic nature of turbulence, not fully representative.
Thus, several realizations for each turbulence intensity have been
computed. The number of independent wingbeats available for averaging
varies between 16 for the lowest and 108 for the largest value of
$Tu$.

The main result of \citep{EKSLS15} was that the ensemble-averaged
forces, torques and the aerodynamic power did not differ from the
values in the laminar case, though the values fluctuated of course.
It was concluded that even in the strongest background turbulence,
no systematic destruction of the leading edge vortex occured, as this
would have resulted in a significant change in the aerodynamic quantities.

\begin{figure}
\begin{centering}
\includegraphics[width=1\textwidth]{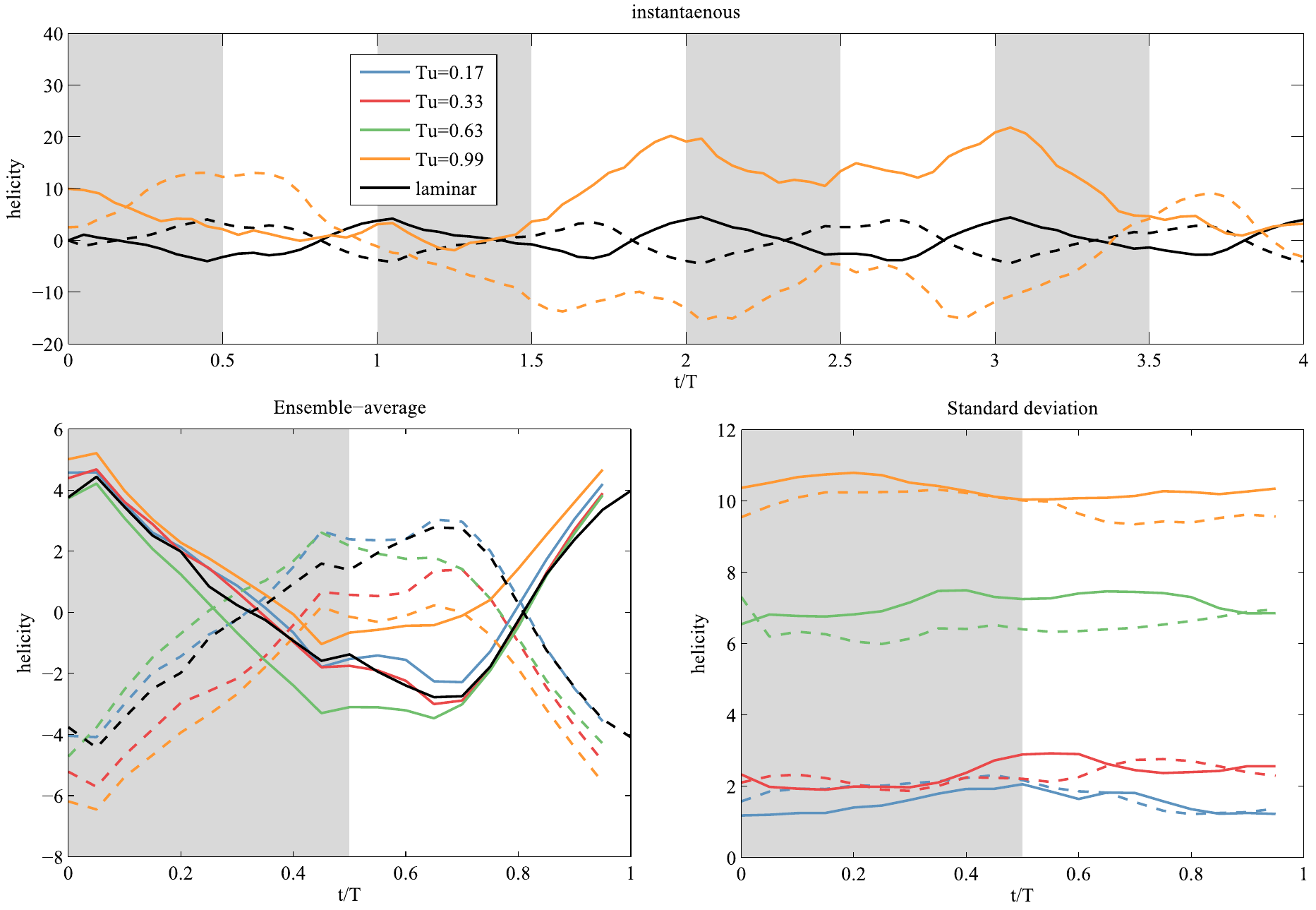}
\par\end{centering}
\caption{Bumblebee for laminar inflow and turbulent inflow with turbulence
intensity $Tu=0.17,0.33,0.63$ and $0.99$. Top: time evolution of
the instantaneous helicity $H$ integrated over the left and right
domain with respect to the vertical center plane of the bumblebee.
Bottom: time evolution of the ensemble averaged helicity $H$ integrated
over the left and right domain with respect to the vertical center
plane of the bumblebee for laminar inflow and turbulent inflow (left)
and corresponding standard deviation (right).}
\label{fig_int_helicity} 
\end{figure}

In the present work, the emphasis lies on the helicity, which we did
not consider previously. Integrating the helicity over the half-space
of the computational domain with respect to the bilateral symmetry
plane of the insect, one obtains the mean helicity generated by the
left and the right wing. The top part of Fig. \ref{fig_int_helicity}
shows the left- and right wing contribution for two individual realizations.
The black line corresponds to laminar inflow, and the integral helicity
is symmetric except for the sign. Their sum is therefore zero, meaning
that the bumblebee produces no net helicity in the wake. By contrast,
the orange line corresponds to a single realization of $Tu=0.99$.
The strong inflow turbulence breaks the symmetry, and thus, even though
the HIT fields do not contain a net helicity, the left- and right
wing contributions do not add to zero. 

The bottom part of Fig. \ref{fig_int_helicity} shows time evolutions
of the ensemble-averaged values of left- and right helicity. The black
line again corresponds to the laminar inflow. The values in the turbulent
simulations however are similar to the laminar ones. This finding
is consistent with \citep{EKSLS15}. The standard-deviation of the
helicity grows with $Tu$ increasing, thus higher $Tu$ implies, as
expected, larger fluctuations.

\section{Conclusion}

By means of high resolution direct numerical simulations we studied
two flow configurations relevant to insect flight. First, a rotating
bumblebee wing at two Reynolds numbers has been considered as canonical
problem, then we passed to a compete bumblebee model, in order to
check if the results obtained in the former can be extrapolated to
the latter.

The revolving wing has been considered at two Reynolds numbers, based
on the wingtip velocity, of about 2000 and 200. A leading edge and
tip vortex is observed in both cases. We found that helicity is not
produced near the leading edge, but instead at a position towards
the trailing edge, and that it is due to the axial flow generated
by the pressure deficit at the wing tip. This flow does not develop
immediately at the leading edge, hence the lack of helicity there.
The vortex core is highly helical, with large values of $H$, and
$h$ near unity, corresponding thus to alignment or anti-alignment
of velocity and vorticity. The nonlinear term is therefore depleted,
and the leading edge vortex can be interpreted as a coherent structure
as proposed in the literature. This finding is important as it provides
a complementary point of view on the observed stability of the LEV,
not in contradiction to other concepts like the axial transport of
excess vorticity \citep{Maxw2007}. An analysis with orthogonal wavelets
allowed us to characterize the most helical scale and its spatial
intermittency. We showed, in agreement with experimental results,
that the integral helicity on the top side of the wing is sensitive
to the Reynolds number, and exhibits, at the higher $Re$ considered,
a significant drop that can be interpreted as vortex bursting. The
aerodynamic force production was indeed not affected by this bursting,
and the burst LEV remained attached to the wing.

We verified then, using the bumblebee model, that these results can
be extrapolated to real insects with their more complex flapping motion,
as opposed to the simple, continuous rotation. Similar features in
the flow were found, namely helical LEVs and tip vortices with opposite
helicity. In addition, turbulent inflow has been imposed, and we confirmed,
in agreement with \citep{EKSLS15}, that turbulence does not alter
ensemble-averaged flight characteristics, also regarding the helicity.

\bigskip{}

\noindent \textbf{\small{}Acknowledgements.}{\small{} Financial support
from the ANR (Grant 15-CE40-0019) and DFG (Grant SE 824\textbackslash{}26
-1), project AIFIT, is gratefully acknowledged and CPU time from the
supercomputer center Idris in Orsay, project i20152a1664. For this
work we were also granted access to the HPC resources of Aix-Marseille
Université financed by the project Equip@Meso (ANR-10-EQPX-29-01).
TE, KS, MF, FL and JS thankfully acknowledge financial support granted
by the ministères des Affaires étrangères et du développement International
(MAEDI) et de l'Education nationale de l'Enseignement supérieur et
de la Recherche (MENESR), and the Deutscher Akademischer Austauschdienst
(DAAD) within the French-German Procope project FIFIT. 
DK gratefully acknowledges the financial support from the JSPS (Japan
Society for the Promotion of Science) Postdoctoral Fellowship, JSPS
KAKENHI Grant Number 15F15061. }{\small \par}

\end{document}